\pdfoutput=1  
\documentclass[sigconf,nonacm,9pt]{acmart}
\microtypesetup{expansion=false}

\usepackage[english]{babel}
\usepackage{enumitem}
\usepackage{booktabs}
\usepackage{multirow}
\usepackage{xcolor}
\usepackage{listings}
\usepackage{subcaption}
\usepackage{wrapfig}
\usepackage{tikz}
\newcommand{\rev}[1]{#1}

\lstdefinestyle{tiara}{
  basicstyle=\ttfamily\footnotesize,
  keywordstyle=\bfseries,
  commentstyle=\itshape\color{gray},
  numbers=none,
  frame=single,
  xleftmargin=4pt,
  xrightmargin=4pt,
  aboveskip=4pt,
  belowskip=4pt,
  columns=fullflexible,
  keepspaces=true,
  morekeywords={Operator,ComputeOp,Load,Store,Memcpy,CAS,CAA,Jump,Loop,Wait,Ret},
  morecomment=[l]{//},
}

\begin{document}

\title{Tiara: A Programmable Line-Rate ISA for Remote Memory Access}

\author{Bojie Li}
\affiliation{%
  \institution{Pine AI}
  \country{}}

\begin{abstract}
RDMA one-sided verbs are the natural primitive for memory disaggregation,
but they require the client to supply the exact remote address.
The 1-RTT performance breaks down when the target address
depends on data that must first be read from remote memory, a pattern we call
the \emph{Indirection Wall}.  Indirection is pervasive: graph traversals
follow pointers hop by hop, address translation walks multi-level page tables,
distributed coordination requires conditional multi-host logic, and
disaggregated LLM inference must resolve paged KV caches through block-table
lookups.  Each level of indirection costs one sequentially dependent network
round-trip, yet offloading to existing RDMA NICs either consumes remote
CPU cycles or has limited throughput.
We present Tiara, a compact, statically verifiable instruction
set that executes on the memory-side NIC\@.  Tiara operators are
pre-registered programs, analogous to eBPF programs in the kernel, that
resolve indirection locally, collapsing multi-RTT dependent chains into a
single round-trip.  On an FPGA-based prototype, Tiara reduces
10-hop graph-traversal latency by \rev{2.85$\times$} over one-sided RDMA while
sustaining \rev{3.4$\times$} higher throughput, cuts page-table walk latency by
\rev{62\%}, \rev{reduces uncontended distributed-lock latency by 2.9$\times$,
achieves 2.8$\times$ throughput for disaggregated PagedAttention at 8\,KB
blocks, and 1.88$\times$ MoE expert-gather latency at 32~experts.}
\end{abstract}


\maketitle


\section{Introduction}\label{sec:intro}

\rev{Datacenter workloads increasingly demand memory beyond a single server's
capacity, separating compute and memory into network-attached
pools~\cite{aguilera2019designing, AIFM}.  One-sided RDMA verbs are the
natural primitive for this \emph{disaggregated} setting: the requester
specifies a remote address and the remote NIC completes the access directly,
without notifying the remote CPU, a model a decade of systems has
meticulously exploited~\cite{FaRM, HERD, FaSST, DrTM_H, mitchell2013using}.
But every verb requires the client to supply the exact remote address at issue
time.  This works when the address is computable locally (e.g., fixed-size HPC
matrices), but a growing class of workloads derives the address from
\emph{remote data that must be read first}.  The client cannot know the next
address until the current read returns, so each dependent lookup costs a full
round-trip that no batching or prefetching can hide.  We call this sequential
bottleneck the \emph{Indirection Wall}, with three recurring patterns:}

\begin{enumerate}[leftmargin=*,nosep]
\item \textbf{Pointer chasing.}  Graph traversals and linked data structures
  store each successor's address in the current node.  Every hop requires one
  sequential network round-trip.  This pattern arises in social-network
  analysis, knowledge-graph queries, and agent-driven graph-RAG pipelines.

\item \textbf{Multi-level translation.}  Page tables, storage indirection
  layers, and block-table lookups resolve addresses through $k$ levels of
  indirection, each requiring a dependent RTT\@.  This pattern is increasingly
  critical for disaggregated LLM inference: vLLM's
  PagedAttention~\cite{kwon2023pagedattention} scatters KV caches into
  non-contiguous blocks, and the client must read a Block Table to discover
  physical addresses before fetching KV data.

\item \textbf{Conditional multi-host coordination.}  Distributed locks and log
  replication use atomic compare-and-swap (CAS) to acquire state on one node,
  then conditionally propagate to replicas, forming a sequence of dependent operations
  spanning multiple hosts.
\end{enumerate}

\noindent
The cost is severe: latency grows as Depth\,$\times$\,RTT, link utilization
collapses between dependent accesses, and the CPU must orchestrate every
intermediate step, precisely the ``killer microseconds''
regime~\cite{attack-of-the-killer-microseconds} where hardware latency
dominates software optimization.  Yue et al.~\cite{yue2025rtt} quantified this for
disaggregated PagedAttention: fetching KV data for a single LLaMA3-70B
request incurs 160 sequential RTTs, leaving a 200\,Gbps link idle 83\% of the time.

\rev{Existing approaches each fall short: chaining RDMA verbs on the
memory-side NIC~\cite{RedN} resolves indirection in 1~RTT but is
throughput-limited; two-sided RPC consumes CPU cores disaggregated memory
nodes may lack; off-path SmartNIC offloading can \emph{increase}
latency~(\S\ref{sec:bf2}).}

\rev{We address this with Tiara, a compact instruction set for the memory-side
NIC\@.  A Tiara \emph{operator} is a pre-registered program, analogous to an
eBPF program in the kernel~\cite{ebpf}.  Its core mechanism is simple:
a \texttt{Load} returns a value into a register that can immediately serve as
the address for the next \texttt{Load}, chaining dependent dereferences with
no round-trips.  Around this core sit control-flow instructions (forward-only
\texttt{Jump}, bounded \texttt{Loop}, async \texttt{Wait}) and integer
\texttt{Compute} for address arithmetic.  Static verification at registration
guarantees bounded execution and sandboxed access.  Tiara attempts no general
computation, only the minimal address resolution that turns an indirect
access into a direct one.}

We make two contributions:
\begin{enumerate}[leftmargin=*,nosep]
\rev{\item \textbf{Tiara}, a \emph{minimal, statically verifiable} NIC-side ISA.
  The contribution is not NIC-side programmability itself (RedN, RMC, NAAM,
  and BF-3 DPA all offer that) but the specific subset that
  simultaneously (i)~collapses dependent dereferences in a hardware-native
  sub-$\mu$s path that software-dispatched ARM/RISC-V cores cannot reach
  (Fig.~\ref{fig:crossover}), and (ii)~admits eBPF-style termination and
  memory-region verification at registration time, so operators are safe
  for multi-tenant sharing without runtime checks.
\item An \textbf{FPGA prototype} on Alveo U50, evaluated on five workloads.
  Compared to one-sided RDMA: 2.85$\times$ lower graph-traversal latency
  at depth~10 and 3.4$\times$ higher throughput at depth~3, 62\% lower
  page-table walk latency, 3.1$\times$ lower distributed-lock latency at
  16~clients, 2.78$\times$ PagedAttention throughput at 8\,KB blocks, and
  1.88$\times$ MoE expert-gather latency at 32~experts.}
\end{enumerate}

Tiara is open source at \url{https://github.com/bojieli/Tiara}.

\section{Motivation}\label{sec:motivation}

\subsection{The Cost of Indirection}\label{sec:cost}

\rev{One-sided RDMA pays one RTT per indirection level; these RTTs are
\emph{sequentially dependent}: the client cannot issue
level-$i{+}1$ until level-$i$ completes, and no batching or pipelining can
help.  Table~\ref{tab:rtt} quantifies this cost across the five workloads we
evaluate, plus sparse attention (NSA), an emerging AI pattern.}

\begin{table}[t]
\centering
\caption{RTT cost of indirection across workloads.}
\label{tab:rtt}
\scalebox{0.8}{%
\begin{tabular}{@{}llcc@{}}
\toprule
\textbf{Workload} & \textbf{Pattern} & \textbf{RDMA} & \textbf{Tiara} \\
\midrule
Graph traversal (depth $d$) & Pointer chase & $d$ RTTs & 1 RTT \\
Page-table walk (3-level) & Multi-level translation & 3+1 RTTs & 1 RTT \\
Dist.\ lock + replication & CAS + cond.\ writes & 5 RTTs & 2 RTTs \\
PagedAttention & Table lookup + gather & 160$^\dagger$ / 2$^\ddagger$ & 1 RTT \\
MoE expert loading & Paged translation & 2 RTTs & 1 RTT \\
\midrule
Sparse attention (NSA) & Score-then-select & 2 RTTs & 1 RTT \\
\bottomrule
\end{tabular}}

{\raggedright\footnotesize
$^\dagger$Unoptimized stop-and-wait, as deployed today~\cite{yue2025rtt}.
$^\ddagger$Optimally batched: 1~RTT to read the block table, then 160
block reads in a 2nd RTT (requires constructing 160 work requests on the client).
\par}
\end{table}

\paragraph{Indirection across the AI stack.}
The Indirection Wall extends beyond classical systems.  In MoE serving with
disaggregated expert weights, the mapping from expert ID to physical location
goes through a translation table, structurally identical to PagedAttention's
Block Table.  DeepSeek's Native Sparse Attention (NSA)~\cite{deepseekv3}
scores compressed keys in remote memory to select which full blocks to
fetch, a ``score-then-select'' pattern where the decision of \emph{what to
read} depends on remote data.  As AI systems adopt disaggregated
memory~\cite{mooncake, distserve, splitwise, cachegen},
data-dependent address resolution becomes a systematic bottleneck.

\paragraph{Indirection vs.\ scatter-gather.}
Not all irregular access patterns are indirection.  In embedding table
lookups~\cite{FlexEMR}, all reads are independent and parallelizable
via RDMA scatter-gather lists.  The Indirection Wall arises specifically when
the address depends on \emph{remote data that must be read first}.

\paragraph{Case Study: Graph Traversal.}
\rev{Social-network graphs stored in disaggregated memory exhibit pure pointer
chasing.  A $k$-hop query (friend-of-friend, influence propagation, community
detection) reads a node, follows an edge to a neighbor, and repeats; each hop's
address depends on the previous hop's data, so no prefetching is possible.
Knowledge graphs and graph-based retrieval-augmented generation (graph-RAG)
pipelines face the same bottleneck: multi-hop reasoning requires sequential
node fetches whose addresses are discovered only at runtime.  With one-sided
RDMA, latency grows as $k \times \text{RTT}$; Tiara resolves all hops via local
DRAM reads, keeping latency near 1~RTT regardless of depth
(\S\ref{sec:eval-graph}).}

\paragraph{Case Study: Page-Table Walk.}
\rev{Consider a compute node accessing remote memory through a 3-level page
table on the memory node.  Translating a virtual address chains three strictly
dependent reads (level-1 $\to$ level-2 $\to$ level-3 entry $\to$ physical
address): one-sided RDMA costs 4~RTTs, while Tiara resolves all three levels
locally in 1~RTT.  This page table is not the OS page table but the
\emph{block-indirection table} disaggregated runtimes build over remote pools
for dynamic reallocation, of which vLLM's Block Table and MoE's expert table
are instances.  ConnectX MTTs translate one flat region but cannot represent
multi-level or application-managed indirection, and IOMMUs operate only on
host-side IO mappings.  This is the on-demand remote-paging path of systems
such as ODRP~\cite{odrp2025}: a node faulting on a remote page must walk a
memory-node-resident translation structure before the fetch, a walk Tiara
collapses into the same round-trip as the fetch.}

\begin{figure}[t]
  \centering
  \includegraphics[width=\columnwidth]{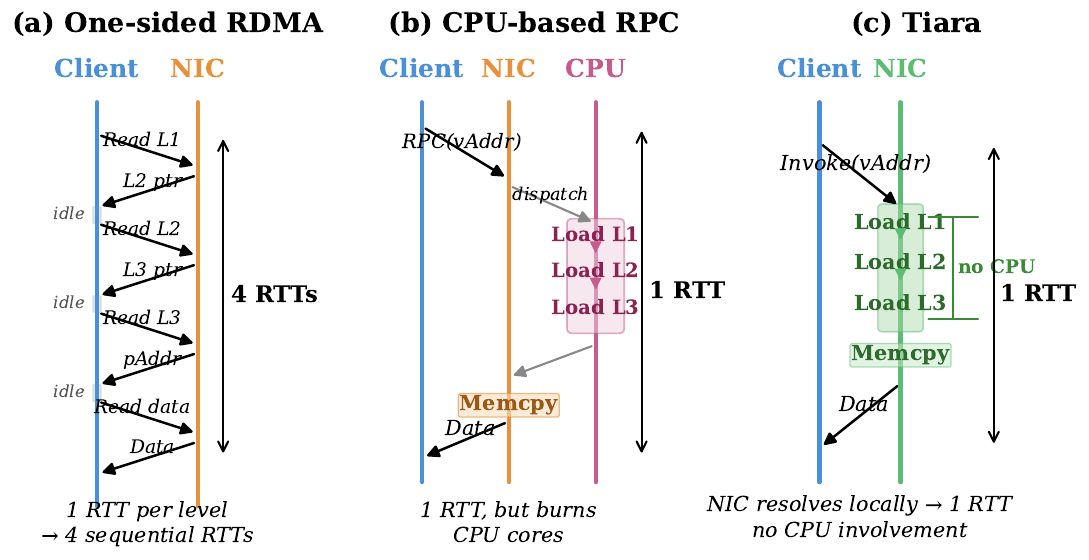}
  \caption{Three approaches to data-dependent remote access (e.g., a 3-level
    page-table walk).
    (a)~One-sided RDMA incurs one RTT per indirection level plus one for
    the final data fetch.  (b)~CPU-based RPC resolves all levels in 1~RTT
    but consumes CPU cores on the memory node.  (c)~Tiara resolves all levels via host
    DRAM on the NIC and returns data in
    1~RTT, without CPU involvement.}
  \label{fig:indirection}
  \vspace{-5pt}
  \Description{Three panels: (a) one-sided RDMA with 4 RTTs for a 3-level page-table walk, (b) CPU-based RPC resolving in 1 RTT with CPU cost, (c) Tiara resolving locally on the NIC in 1 RTT.}
\end{figure}

\begin{figure}[t]
  \centering
    \includegraphics[width=0.65\columnwidth]{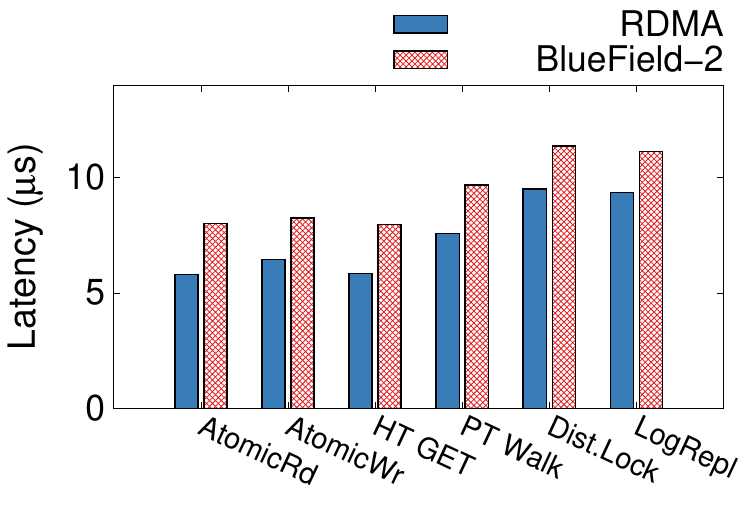}
    \captionof{figure}{BlueField-2 vs.\ one-sided RDMA latency.}
    \label{fig:bf2}
\end{figure}

\begin{figure}[t]
  \centering
  \includegraphics[width=0.8\columnwidth]{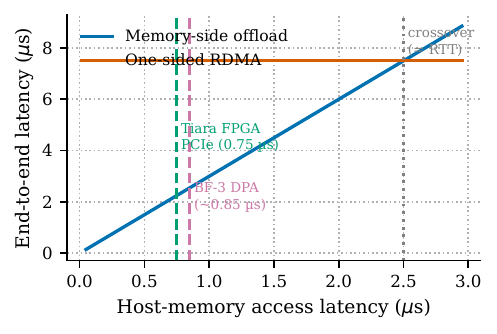}
    \captionof{figure}{Offloading beats RDMA when host-memory
      latency $<$ RTT.}
    \label{fig:crossover}
\end{figure}

\begin{figure}[t]
  \centering
  \resizebox{0.9\columnwidth}{!}{%
    \begin{tikzpicture}[
      block/.style={draw, thick, rounded corners=1.5pt, minimum height=0.5cm,
                    text centered, font=\small, inner sep=3pt},
      mp/.style={draw, thick, fill=yellow!12, minimum width=1.3cm,
                 minimum height=1.0cm, rounded corners=1pt, inner sep=2pt,
                 font=\scriptsize, text centered, align=center},
      >=stealth,
    ]
    \node[block, fill=gray!12, minimum width=2cm] (remote)
      at (5, 7.0) {Remote Nodes};

    \fill[blue!4, rounded corners=3pt] (0, 1.8) rectangle (10.2, 6.4);
    \draw[very thick, rounded corners=3pt] (0, 1.8) rectangle (10.2, 6.4);
    \node[font=\small\bfseries, anchor=north west] at (0.2, 6.3)
      {Tiara NIC};

    \node[block, fill=orange!18, minimum width=9.4cm] (rdma) at (5.1, 5.55)
      {RDMA Engine};

    \node[block, fill=green!12, minimum width=9.4cm] (disp) at (5.1, 4.75)
      {Task Dispatcher};

    \node[mp] (mp0) at (1.8, 3.5)
      {\textbf{MP$_0$}\\Regs, ALU\\IStore};
    \node[mp] (mp1) at (3.8, 3.5)
      {\textbf{MP$_1$}\\Regs, ALU\\IStore};
    \node[font=\normalsize] at (5.8, 3.5) {\ldots};
    \node[mp] (mp7) at (7.8, 3.5)
      {\textbf{MP$_7$}\\Regs, ALU\\IStore};

    \node[block, fill=purple!12, minimum width=9.4cm] (dma) at (5.1, 2.3)
      {PCIe DMA Engine};

    \draw[->, thick] (rdma) -- (disp);
    \draw[->, thick] ([xshift=-8pt]disp.south) -- ++(0,-0.18) -| (mp0.north);
    \draw[->, thick] (disp.south) -- ++(0,-0.18) -| (mp1.north);
    \draw[->, thick] ([xshift=8pt]disp.south) -- ++(0,-0.18) -| (mp7.north);
    \draw[<->, thick] (mp0.south) -- (mp0.south |- dma.north);
    \draw[<->, thick] (mp1.south) -- (mp1.south |- dma.north);
    \draw[<->, thick] (mp7.south) -- (mp7.south |- dma.north);

    \draw[<->, thick, densely dashed, blue!70!black]
      ([yshift=2pt]mp7.north east) -- ++(0.5, 0) |- ([xshift=-3pt]rdma.east)
      node[pos=0.25, right, font=\scriptsize, text=blue!70!black] {Memcpy};

    \draw[<->, thick] (remote.south) -- (rdma.north)
      node[midway, right, font=\scriptsize] {RDMA};

    \draw[<->, very thick] (5.1, 1.8) -- (5.1, 1.45)
      node[midway, right, font=\scriptsize] {PCIe};

    \fill[gray!6, rounded corners=3pt] (0, 0.1) rectangle (10.2, 1.45);
    \draw[thick, rounded corners=3pt] (0, 0.1) rectangle (10.2, 1.45);

    \node[block, fill=cyan!12, minimum width=2.4cm, align=center] (dram)
      at (2.5, 0.75) {Host DRAM\\[-1pt]{\scriptsize(Memory Regions)}};
    \node[block, fill=gray!15, minimum width=2cm] (cpu)
      at (5.1, 0.75) {Host CPU};
    \node[block, fill=red!10, minimum width=2.4cm, align=center] (cv)
      at (7.8, 0.75) {Compiler \&\\[-1pt]Verifier};

    \node[font=\small\bfseries, anchor=north west] at (0.2, 1.35)
      {Host};

    \draw[<->, thick] (dram) -- (cpu);
    \draw[<->, thick] (cpu) -- (cv);

    \draw[->, thick, dashed, red!60!black]
      (cv.north) -- (cv.north |- dma.south)
      node[pos=0.55, right, font=\scriptsize, text=red!60!black]
      {register operators};
    \end{tikzpicture}%
    }
    \captionof{figure}{Tiara NIC architecture.}
    \label{fig:arch}
    \Description{Block diagram: Tiara NIC with RDMA engine, task dispatcher,
      8 memory processors, PCIe DMA engine; host side with DRAM, CPU,
      compiler and verifier.}
\end{figure}
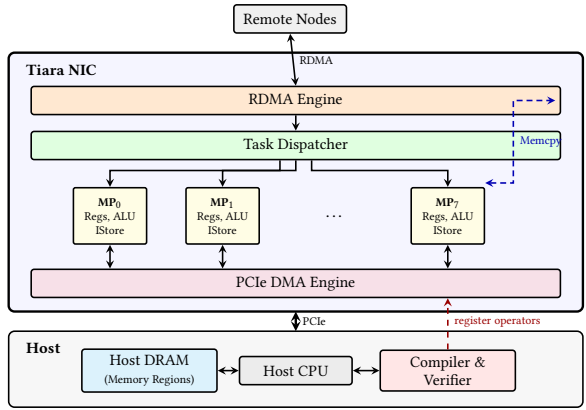

\subsection{Existing Approaches}\label{sec:existing}

\rev{Figure~\ref{fig:indirection} contrasts three approaches.
CPU-based RPC~(b) resolves all levels on the memory node in 1~RTT, but
dedicates CPU cores to address arithmetic.
NIC-side approaches~(c) avoid CPU involvement but, as we show below, each
introduces its own bottleneck.}

\paragraph{Verb chaining.}
\rev{RedN~\cite{RedN} proved RDMA Turing-complete by chaining self-modifying work
requests (WRs, the descriptors that tell the NIC what operation to perform).
RedN executes on the memory-side NIC, resolving indirection in a single
RTT\@.  However, it relies on \emph{doorbell ordering}: the CPU must signal
(``ring a doorbell'' via a PCIe write) after posting each WR, and the NIC
processes them one at a time.  This forces the NIC to fetch each WR
individually from host memory (${\sim}$0.54\,$\mu$s per
WR~\cite{RedN}), limiting throughput to ${\sim}$1\,Mops,
65$\times$ below raw RDMA Read rates~(\S\ref{sec:eval-graph}).}

\paragraph{CPU-based RPC.}
\rev{Two-sided RPC resolves addresses in 1~RTT, but software RPC consumes
22~CPU cores to saturate a 25\,Gbps NIC's message
rate~\cite{eRPC}.  Disaggregated memory nodes increasingly lack
general-purpose CPUs entirely (CXL memory pools, memory
blades)~\cite{aguilera2019designing}.  Even when a CPU is available, RPC
dispatch overhead adds
1 to 3\,$\mu$s~\cite{eRPC, ousterhout2019shenango}, often exceeding the
useful work.}

\paragraph{SmartNIC offloading.}
\rev{Running C/C++ on off-path SmartNIC ARM cores\label{sec:bf2} BlueField
DPU~\cite{DPU} and FlexIO DPA~\cite{flexio} avoid the host CPU but offer no
termination guarantee or memory sandboxing.  Worse, it is not even faster.
To validate this, we implement Tiara operators on NVIDIA
BlueField-2~\cite{Mellanox-BlueField-2, OffPathSmartNIC}, where the ARM
cores access host memory via internal RDMA\@.  Two servers are connected
back-to-back via a DAC cable (RTT~${\sim}$1.9\,$\mu$s), giving the SmartNIC
the shortest possible network RTT\@.
Figure~\ref{fig:bf2} shows that offloading to BlueField-2
\textbf{increases latency for every operator}: an atomic read regresses by
38\%.  The root cause is that each host-memory access costs 1.7\,$\mu$s via
internal RDMA, close to the 1.9\,$\mu$s cable RTT.
Figure~\ref{fig:crossover} sweeps host-memory latency analytically:
the crossover where offloading becomes worthwhile occurs at the network RTT\@.
Tiara's PCIe DMA at 0.75\,$\mu$s places it well below the crossover.}

\paragraph{Active messaging and on-path DPA.}
\rev{NAAM~\cite{naam} runs pre-registered eBPF on host or SmartNIC ARM cores;
BF-3 DPA~\cite{bf3dpa} adds 16 on-path RISC-V cores with $\sim$0.85\,$\mu$s
host-memory access.  Both share Tiara's pre-registration vision and would
collapse multi-RTT chains, and NAAM's eBPF offers richer programmability
(maps, helpers) hard to realize in fixed logic.  Tiara trades that generality
for two properties these designs cannot match.  \emph{Latency}: BF-3 DPA's
0.85\,$\mu$s host access nearly matches Tiara's 0.75\,$\mu$s, but Tiara's
hardware MPs commit a load one cycle after writeback with no dispatch, while
off-path BF-2 ARM cores regress (Fig.~\ref{fig:bf2}).  BF-3 DPA's 0.85\,$\mu$s
stays below the RTT crossover (Fig.~\ref{fig:crossover}), a viable target that
narrows the gap.  One could instead restrict the source language and target an
existing ISA (Tiara itself compiles from a restricted OpenCL~C subset,
\S\ref{sec:compiler}), but that buys static safety on \emph{any} target, not
the sub-$\mu$s per-hop cost.  The minimal ISA is what
makes this cheap: a 2.95\,K-LUT MP (\S\ref{sec:setup}) replicates 8$\times$ in
${\sim}$3\% of a ConnectX die, whereas a RISC-V/ARM core is one-to-two orders
larger and cannot be replicated at line rate; the contribution is this
hardware/ISA co-design, not verification alone.  \emph{Safety}: NAAM leans on
eBPF's runtime verifier, while Tiara proves termination and region isolation
statically with no runtime guard.  Stateless data planes (P4~\cite{p4lang},
NPUs such as Netronome) cannot express data-dependent loops or dependent loads
against host memory, so none of the workloads in Table~\ref{tab:rtt} fit.}

\section{Tiara Design}\label{sec:design}

\rev{Figure~\ref{fig:arch} shows the overall architecture.  The Tiara NIC
contains an RDMA engine, a task dispatcher, and 8 lightweight memory
processors (MPs), each with a register file (16$\times$64\,b), an integer
ALU, a depth-8 loop stack, a 32-entry in-flight async counter, and a BRAM
instruction store for pre-registered operators.
At registration time (dashed path), the host-side compiler and static
verifier check operators before loading them into the NIC's instruction
stores via PCIe.  At execution time (solid paths), incoming RDMA requests
are dispatched to an MP, which accesses host DRAM via PCIe DMA.
We detail the programming model~(\S\ref{sec:progmodel}), instruction
set~(\S\ref{sec:instructions}), and
compiler and verifier~(\S\ref{sec:compiler}) below.}

\paragraph{Execution and multi-tenant operation.}
\rev{An MP is a sequential scalar core (11-state FSM, no cache, no branch
prediction, no out-of-order): register-chained loads are made correct by
stalling fetch until writeback.  A 256-entry $\textit{op\_id}\!\rightarrow\!
\textit{start\_pc}$ table in front of the dispatcher routes incoming
requests to any registered operator in O(1), so one NIC hosts many tenants'
operators concurrently, an opaque per-task tag routing responses to the right
caller.  Isolation is \emph{static}: every operator is verified at
registration to access only its declared regions, so the runtime needs no
per-access check and one tenant's operator cannot reach another tenant's
memory regardless of how it executes.  The memory subsystem is uncached and
region-partitioned (a
device-id router sends local accesses to PCIe DMA, remote ones to the RDMA
engine); we add no MP-side cache, since indirection's poor locality offers
nothing to cache.  The shared resources are the 8 MPs and the host PCIe
channel; the current dispatcher is work-conserving but not weighted, and
per-tenant token-bucket scheduling is a one-state extension
left to future work.  Higher line rates scale by instantiating more MPs, each
only ${\sim}$3\,K~LUT (\S\ref{sec:setup}).}

\subsection{Programming Model}\label{sec:progmodel}

\rev{An \textbf{operator} is a small program of Tiara instructions, registered
on a NIC before use (analogous to loading an eBPF program into the kernel);
registration triggers compilation and static verification~(\S\ref{sec:compiler}).
A client invokes it by sending a single message with the operator~ID,
parameters (up to 8 registers), and optional inline data.  The NIC creates a
\textbf{task} backed by a register file (16$\times$64\,b) and executes the
operator without involving the host CPU\@.}

Pre-registration is key: only the operator~ID and parameters travel on the
wire: one message per invocation.  In contrast, PRISM~\cite{PRISM} and
RedN~\cite{RedN} transmit multiple work requests per task on every call,
consuming PCIe and network bandwidth proportional to task complexity.

An operator accesses local memory via DMA and remote memory on \emph{any} node
via \texttt{Memcpy} with unified addressing, enabling multi-host orchestration
in a single invocation, e.g., a distributed lock operator CAS-acquires a
latch, replicates state to two replicas via parallel async \texttt{Memcpy},
waits for acknowledgments, and returns, all without CPU involvement on any
node.  PRISM's chaining primitives are scoped to a single client-server pair;
Tiara's unified addressing removes this limitation.

\subsection{Instruction Set}\label{sec:instructions}

Table~\ref{tab:instructions} lists Tiara's instructions.  The design principle
is \emph{minimal but sufficient}: enough to express all indirection patterns in
Table~\ref{tab:rtt}, simple enough to verify statically and implement in
hardware.  We derived this set by decomposing all workloads in
Table~\ref{tab:rtt} into elemental operations: every instruction is required
by at least one workload, and removing any breaks that workload.

\begin{table}[t]
\centering
\small
\caption{Tiara instruction set.}
\label{tab:instructions}
\begin{tabular}{@{}lp{6.2cm}@{}}
\toprule
\textbf{Instruction} & \textbf{Description} \\
\midrule
\texttt{Load/Store} & Register $\leftrightarrow$ local memory.
  A loaded value can be the next address. \\
\texttt{Memcpy} & Bulk transfer with unified (device, addr) addressing;
  subsumes RDMA Read/Write. \\
\texttt{CAS/CAA} & Atomic compare-and-swap / compare-and-add. \\
\midrule
\texttt{Jump} & Forward-only conditional branch. \\
\texttt{Loop(M,N)} & Execute next $N$ ops for $M$ iterations. \\
\texttt{Wait} & Block until in-flight async ops $\leq$ threshold. \\
\texttt{Ret} & Return result to caller. \\
\midrule
\texttt{ComputeOp} & Integer arithmetic, logical, shift for address computation. \\
\bottomrule
\end{tabular}
\end{table}

\rev{Three design choices merit emphasis.  The key enabler is
\textbf{register-chained loads}: a \texttt{Load} writes its result into a
register that a subsequent \texttt{Load} can use as its address operand the
very next cycle, turning a multi-RTT pointer chase into local memory accesses,
each ${\sim}$0.75\,$\mu$s on our FPGA prototype versus a full network RTT\@.}

\rev{Second, \textbf{unified addressing}: addresses are (host\_id, region\_id,
offset) tuples.  A \texttt{Memcpy} with remote source and local destination
is an RDMA Read; with remote destination, an RDMA Write.  This eliminates
separate verb types and simplifies multi-host operators.}

\rev{Third, \textbf{async~+~wait}: \texttt{Memcpy} instructions execute
asynchronously; \texttt{Wait(threshold)} synchronizes.  Setting the threshold
to~0 waits for all in-flight operations; a threshold~$>0$ enables
quorum-style synchronization (e.g., proceed when all but one replica
acknowledges).  It also enables pipelining: issuing KV block reads as
block-table entries are resolved, overlapping address resolution with
transfer.  An async op to a failed node times out and sets an error flag the
operator can test via conditional \texttt{Jump} to execute a fallback path
(e.g., skip the failed replica) or return an error to the caller.}

\subsection{Compiler and Verifier}\label{sec:compiler}

\paragraph{Compiler.}
\rev{Operators are written in a restricted subset of OpenCL
C~\cite{munshi2009opencl} so that every operator's control flow is a Static
Control Part (SCoP)~\cite{girbal2006semi}, making termination and resource
bounds decidable at compile time.  An LLVM-based~\cite{lattner2008llvm}
toolchain lowers OpenCL C to IR (standard passes run unmodified); a custom
Tiara back-end does linear-scan register allocation~\cite{poletto1999linear},
flattens loops into bounded \texttt{Loop(M,N)}, selects opcodes, and inlines
write data with the request to fold a 2-RTT read-then-write into 1~RTT\@.  The
resulting binaries are compact (typically 10 to 50 instructions).}

\paragraph{Static verification.}
\rev{Operators are verified at registration time, before they ever execute on
the data path.  This is the critical differentiator from SmartNIC C
programming:}

\begin{enumerate}[leftmargin=*,nosep]
\item \textbf{Termination guarantee.}  Forward-only jumps (no backward
  branches) and bounded loop iterations give every operator a statically
  computable upper bound on execution steps.
  The verifier computes this bound and rejects
  operators exceeding a configurable limit.

\rev{\item \textbf{Fine-grained access control.}  One-sided RDMA exposes entire
  memory regions to any client with the region key.  Tiara's verifier checks
  at registration that every memory access falls within server-configured
  regions.  Server-registered operators can encapsulate restrictive
  logic (e.g., return usernames but never passwords), and
  the client never sees raw memory, making Tiara both
  \emph{more secure} and \emph{more flexible} than one-sided RDMA.}
\end{enumerate}

\begin{figure}[t]
  \begin{lstlisting}[style=tiara]
  Operator DistLock(latch, state, newVal,
                    r1Dev, r1Addr, r2Dev, r2Addr):
    Loop(MAX_RETRIES, 2)         // bounded CAS retry
      ok = CAS(latch, 0, 1)      // local memory
      Jump(ok == 0, acquired)
    Ret(FAIL)
    acquired:
      old = Load(state)
      Store(state, newVal)
      Memcpy((r1Dev,r1Addr), state, 8)  // async
      Memcpy((r2Dev,r2Addr), state, 8)  // async
      Wait(0)                     // both replicas ACK
      Store(latch, 0)             // release
      Ret(old)
  \end{lstlisting}
  \caption{Distributed lock operator.}
  \label{fig:operator}
\end{figure}

\subsection{Example Operators}\label{sec:example}

\paragraph{Distributed lock.}
\rev{Figure~\ref{fig:operator} shows a lock operator spanning three hosts: the
NIC acquires a latch via local \texttt{CAS} (retried in a bounded
\texttt{Loop}), updates state, replicates to two backups via parallel async
\texttt{Memcpy}, and releases, collapsing RDMA's 5~sequential RTTs into~2
(client$\to$primary, then primary$\to$replicas in parallel) with no CPU
involved.}

\paragraph{Page-table walk.}
Three chained \texttt{Load}s (each using the previous result as the next
address) resolve a 3-level page table; a final \texttt{Memcpy} transfers the
data page.  RDMA: 4~RTTs; Tiara: 1~RTT\@.  The same pattern extends to
PagedAttention: a \texttt{Loop} iterates over block~IDs, resolving each via
\texttt{Load} and issuing async \texttt{Memcpy} for the KV block.

\section{Evaluation}\label{sec:eval}

\subsection{Setup}\label{sec:setup}

\paragraph{Testbed.}
\rev{Two dual-socket Intel Xeon Gold 6326 servers are connected via 100\,GbE
QSFP28 through a single TOR switch (RDMA Read RTT ${\sim}$2.5\,$\mu$s).
Baselines run on NVIDIA BlueField-2 NICs; for Tiara the memory-side NIC is an
AMD Alveo U50 FPGA running the Corundum~\cite{corundum} NIC stack, accessing
host DRAM via PCIe\@.}

\paragraph{Implementation.}
\rev{We implement Tiara's execution engine as an extension to the Corundum NIC
pipeline on the Alveo U50: 8 memory processors at 200\,MHz, each with a
16-register file and a 1024-entry BRAM instruction store.  Each MP is only
2.95\,K~LUT (+1.69\,K~FF); the single-MP build meets timing at 200\,MHz
post-route with $+0.184$\,ns slack (Fmax ${\sim}$208\,MHz), matching the
Corundum app clock.  The remaining ${\sim}$24\,K~LUT is the OOC memory stub,
replaced by Corundum DMA in deployment.}

\paragraph{Measurement methodology.}
\rev{Tiara latencies are cycle-accurate on the Verilator model (5\,ns clock,
150-cycle PCIe DMA, 500-cycle RDMA RTT, calibrated to the U50 build);
saturated throughput is derived from measured latency for 8 MPs $\times$ 12
outstanding tasks (96 dispatcher slots).  Non-Tiara baselines are analytical
models, parameterized identically and printed alongside the measured data in
each \texttt{.dat} file: RDMA~\cite{RDMA} at 2.5\,$\mu$s RTT; RPC
(eRPC-style~\cite{eRPC}) at 1.5\,$\mu$s dispatch + 0.17\,$\mu$s/cached-DRAM
hop, at 16 and 22 (saturation) cores; RedN~\cite{RedN} at 1.1\,$\mu$s/WR;
PRISM~\cite{PRISM} at 0.5\,$\mu$s/hop (graph only; it lacks arithmetic,
loops, multi-host coordination).  BF-2 points (Fig.~\ref{fig:bf2}) are measured
on a physical NIC; the BF-3 DPA marker (Fig.~\ref{fig:crossover}) is from
NVIDIA's datasheet~\cite{bf3dpa}.}

\begin{figure}[t]
  \centering
  \includegraphics[width=0.9\columnwidth]{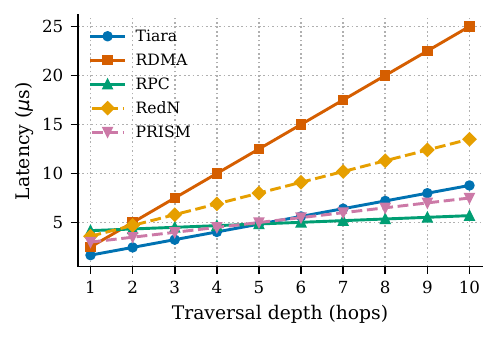}
  \captionof{figure}{Graph traversal latency vs.\ depth.}
  \label{fig:graph}

  \includegraphics[width=0.9\columnwidth]{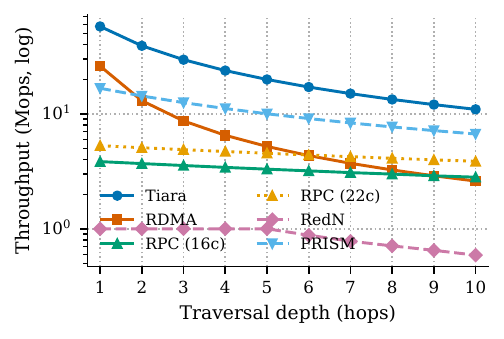}
  \captionof{figure}{Graph traversal throughput vs.\ depth.}
  \label{fig:graph-tput}
  \vspace{-5pt}
\end{figure}

\begin{figure*}[t]
  \centering
  \begin{minipage}[t]{0.33\textwidth}
    \centering
    \includegraphics[width=\textwidth]{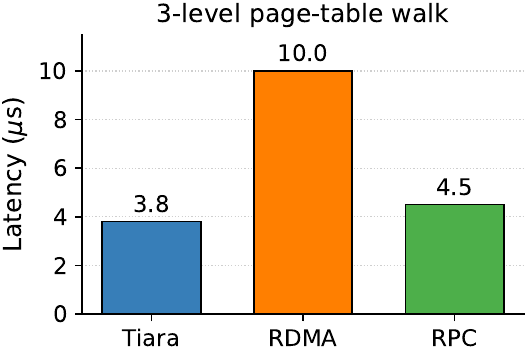}
    \captionof{figure}{Page-table walk latency.}
    \label{fig:pt}
  \end{minipage}\hfill
  \begin{minipage}[t]{0.33\textwidth}
    \centering
    \includegraphics[width=\textwidth]{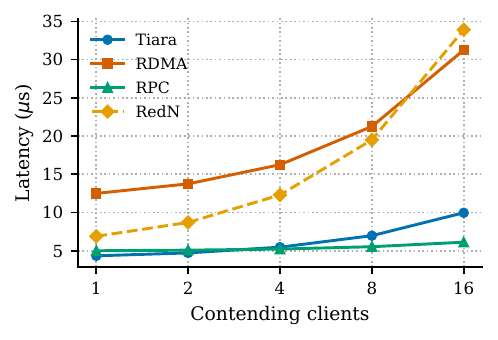}
    \captionof{figure}{Distributed lock latency vs.\ contention.}
    \label{fig:lock}
  \end{minipage}\hfill
  \begin{minipage}[t]{0.33\textwidth}
    \centering
    \includegraphics[width=\textwidth]{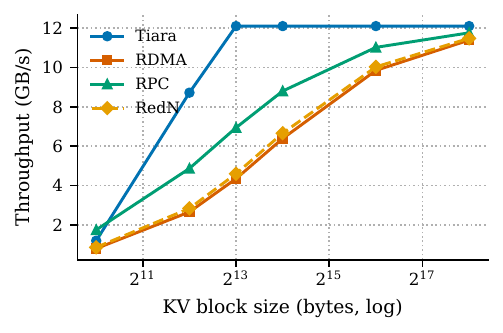}
    \captionof{figure}{PagedAttention throughput vs.\ KV block size.}
    \label{fig:paged}
  \end{minipage}
  \vspace{-10pt}
\end{figure*}

\subsection{Graph Traversal}\label{sec:eval-graph}

\paragraph{Setup.}
A social network graph in remote RDMA-registered memory; 64-byte nodes with
adjacency-list pointers.  The Tiara operator performs a depth-limited walk.

\paragraph{Results.}
\rev{Figure~\ref{fig:graph} shows latency vs.\ depth.  RDMA grows linearly at
$d \times \text{RTT}$.  RPC is nearly flat (all hops resolve via node-local
DRAM, ${\sim}$0.17\,$\mu$s/hop) but dispatch overhead dominates and it consumes
a CPU core per traversal.  Tiara scales at 1\,RTT\,+\,$d \times 0.79$\,$\mu$s,
2.85$\times$ faster than RDMA at depth~10 (8.78 vs.\ 25.0\,$\mu$s).  RedN and
PRISM also run on the memory-side NIC; PRISM scales at ${\sim}$0.5\,$\mu$s/hop,
RedN higher from doorbell ordering~(\S\ref{sec:motivation}).  In a production
ASIC, Tiara's per-hop cost would match PRISM's with a richer ISA; the current
gap is inherent to FPGA PCIe latency~\cite{kv_direct}.}

\paragraph{Latency vs.\ RPC.}
\rev{Tiara's per-hop cost is 0.79\,$\mu$s (PCIe to host DRAM) vs.\ 0.17\,$\mu$s
for cached-DRAM RPC, so RPC overtakes Tiara on latency beyond depth~5.  Tiara
still wins where it matters: saturated throughput (below), preserving (or
eliminating) the memory node's CPU (the CXL/memory-blade case RPC cannot
serve), and footprint (8 MPs in ${\sim}$3\% of a ConnectX die vs.\ $\geq$22
cores per 25\,GbE NIC for saturated RPC).}

\paragraph{Throughput.}
\rev{Figure~\ref{fig:graph-tput} shows saturated throughput (log scale).
Tiara reaches 29.5\,Mops at $d{=}3$, 6.1$\times$ higher than RPC even
at its 22-core saturation point (4.88\,Mops), and 8.3$\times$ higher than
RPC at the 16-core paper baseline (3.55\,Mops).
PRISM tracks close to RDMA (NIC-native primitives, no doorbell ordering).
\emph{RedN sustains only ${\sim}$1\,Mops, ${\sim}$26$\times$ below RDMA at
depth~1}, due to doorbell-ordering overhead~(\S\ref{sec:motivation}),
serializing execution across only 8~processing units.
Tiara avoids this: operators reside in FPGA fabric with no WR fetching.}

\subsection{Page-Table Walk}\label{sec:eval-pt}

\paragraph{Setup.}
A 3-level page table in remote RDMA-registered memory (8-byte entries, table
sizes representative of a 256\,GB disaggregated memory pool).

\paragraph{Results.}
\rev{Figure~\ref{fig:pt} shows latency.  RDMA needs 4~RTTs (10.0\,$\mu$s);
Tiara resolves all three levels via PCIe in 1~RTT (3.75\,$\mu$s), a 62\%
reduction (2.7$\times$).  RedN needs extra WRs for shift/mask arithmetic per
level, amplifying its doorbell overhead.  On throughput Tiara sustains
${\sim}$25\,Mops vs.\ RDMA's 0.1\,Mops, since each translation is one network
message, not four.}

\subsection{Distributed Lock}\label{sec:eval-lock}

\paragraph{Setup.}
A read-write lock replicated on one primary and two replicas.  Lock
acquisition: CAS to acquire latch $\to$ read/update lock state $\to$ replicate
to 2 replicas $\to$ release latch.

\paragraph{Results.}
\rev{Figure~\ref{fig:lock} shows lock-acquire latency under 1 to 16 contending
clients.
Without contention, RDMA requires 5 sequential RTTs (CAS + read + 2~replica
writes + release); Tiara collapses this to 2~RTTs: the first RTT delivers
the request to the primary NIC, which executes the CAS locally and issues
parallel replica writes via async \texttt{Memcpy}; the second RTT covers
the replica acknowledgments (\texttt{Wait}).  No host CPU is involved on any
node.
RedN reduces RDMA's 5~RTTs to 1~RTT but pays doorbell-ordering
overhead for the replica write WR chains.
Under contention (16~clients), RPC degrades the least
(${\sim}$1.2$\times$) because CAS retries are nanosecond-scale
CPU-local operations, overtaking Tiara at ${\sim}$4~clients, the same
axis as the graph-traversal latency tradeoff above: RPC wins on per-client
latency when a CPU core is available, Tiara wins when the memory node has no
CPU to dedicate or when concurrent throughput matters more.
RedN degrades ${\sim}$4.9$\times$ from 1 to 16 clients (doorbell overhead
per retry), and RDMA 2.5$\times$ (each failed CAS incurs an RTT).}

\subsection{MoE Expert Gather}\label{sec:eval-moe}
\rev{A MoE serving layer fetches $k$ expert-weight slabs (8\,KB each) through a
translation table indexed by selected expert IDs, structurally the same
block-table indirection as PagedAttention.  Tiara is cycle-accurate on the
prototype; RDMA/RPC use the \S\ref{sec:setup} baselines.  At 32~experts: Tiara
14.2\,$\mu$s, RDMA 26.7\,$\mu$s (1.88$\times$), RPC 41.7\,$\mu$s
(2.93$\times$); the gap grows with $k$ as RPC's per-expert dispatch dominates
while Tiara stays pipelined via async \texttt{Memcpy}.}

\subsection{Disaggregated PagedAttention}\label{sec:eval-paged}

\paragraph{Setup.}
A memory node holds a vLLM-style paged KV cache~\cite{kwon2023pagedattention}.
A Block Table maps logical block~IDs to physical addresses; task: fetch
8\,MB of KV data (one layer's KV cache for 2048 tokens, LLaMA3-70B) over 100\,GbE\@.
We vary the KV block size and measure effective throughput (GB/s),
which captures both per-block overhead and data transfer cost.
The Tiara operator resolves each block via \texttt{Load}
and pipelines resolution with transfer via async \texttt{Memcpy}.
We compare against optimally batched RDMA (2~RTTs), RPC, and RedN\@.

\paragraph{Results.}
\rev{Figure~\ref{fig:paged} shows throughput vs.\ block size.  At small blocks
(1 to 4\,KB) per-block overhead dominates: Tiara reaches 8.7\,GB/s at 4\,KB while
batched RDMA achieves only 2.7\,GB/s (client-side WR construction).  Tiara
saturates effective line rate (${\sim}$12\,GB/s) at just 8\,KB blocks
(2.8$\times$ batched RDMA), because its pipelined resolve-then-transfer hides
resolution behind transfer; other systems converge only at ${\geq}$256\,KB.
RedN tracks RPC at large blocks but lags at small ones (per-block doorbell
ordering).}

\section{Related Work}\label{sec:related}

\paragraph{Programmable RDMA and active messaging.}
\rev{RedN~\cite{RedN} chains self-modifying work requests in 1~RTT but at
limited throughput~(\S\ref{sec:motivation}); PRISM~\cite{PRISM} adds
NIC-native indirection and chaining; Tiara stores operators in NIC BRAM,
avoiding per-WR PCIe fetches.  RMC~\cite{RMC} and NAAM~\cite{naam} share
Tiara's pre-registration model (software cores running C or eBPF); the
differentiator is the hardware MP path that drops per-hop cost below the
software-dispatch floor (Fig.~\ref{fig:crossover}).  ODRP~\cite{odrp2025},
Storm~\cite{Storm}, and KV-Direct~\cite{kv_direct} achieve 1-RTT with
data-structure-specific paths; Tiara supports many patterns from one design.}

\paragraph{SmartNIC, disaggregated memory, and in-network computing.}
\rev{FlexIO DPA~\cite{flexio} and BlueField DPU~\cite{DPU, Mellanox-BlueField-2}
offer general-purpose C; Tiara trades generality for static guarantees.
ClickNP~\cite{ClickNP}, iPipe~\cite{iPipe}, Floem~\cite{Floem},
AccelTCP~\cite{AccelTCP}, and StRoM~\cite{StRoM} process at the packet and
transport level.  FaRM~\cite{dragojevic2015no}, DrTM+R~\cite{DrTM_R},
FORD~\cite{FORD}, Pilaf~\cite{mitchell2013using}, and AIFM~\cite{AIFM} minimize
indirection via RDMA-friendly layouts; Tiara handles the residual at the NIC.
NetLock~\cite{NetLock}, NetCache~\cite{NetCache}, 1Pipe~\cite{1pipe},
P4~\cite{p4lang}, and nanoPU~\cite{nanoPU} target switch-side offload; Tiara is
complementary on the memory-side NIC.}

\section{Conclusion}\label{sec:conclusion}

\rev{The Indirection Wall, where the remote address depends on remote data, is
a fundamental obstacle for one-sided RDMA across disaggregated memory, graph
analytics, and LLM inference, and CXL pooling that removes host CPUs will only
widen it.  Tiara shows that a compact, statically verifiable NIC-side ISA
collapses multi-RTT indirection into a single round-trip.  Extending the
integer-only ISA to floating-point workloads without breaking static verification is future work.}

\section{Acknowledgments}

This paper was initially submitted to APNet 2026 and accepted for publication.
We thank the anonymous reviewers and our shepherd for their valuable feedback.
We used Cursor and Claude Code extensively for code generation and paper
writing. Because ACM and APNet 2026 policies do not accept papers produced substantially by
generative AI, we decided to withdraw the paper from APNet 2026 and publish it
on arXiv instead.

This work was initially a submission to SIGCOMM 2023 during the author's time
at Huawei. The current implementation is clean-slate and does
not use any proprietary materials from Huawei. We thank the collaborators at
Huawei who gave rise to the initial idea and the initial experiments for this
paper. The author list of that earlier version was: Chengjun Jia (Tsinghua
University), Bojie Li, Lijun Li, Xiaoping Fan, Changhu Chen, Haifeng Lin,
Haonan Chen, Yong Chao, Uri Hasson, Eti Siminchi, Zvika Rubin, Shamir
Rabinovitch, Mingxiang Li, Han Ruan, Kai Zheng, Jingbin Zhou, and Kun Tan (all
Huawei).

\bibliographystyle{ACM-Reference-Format}
\bibliography{reference}

\end{document}